\documentclass[preprint,showpacs,preprintnumbers,amsmath,amssymb]{revtex4}
\usepackage{graphicx}
\usepackage{dcolumn}
\usepackage{bm}
\usepackage{mathrsfs}
\usepackage{amsmath,amssymb,graphics}

\begin{document}

\title{Entropy/Area spectra of the charged black hole from quasinormal modes}

\author{Shao-Wen Wei\footnote{weishaow06@lzu.cn},
        Yu-Xiao Liu\footnote{liuyx@lzu.edu.cn, corresponding author.},
        Ke Yang\footnote{yangke09@lzu.cn},
        Yuan Zhong\footnote{zhongy2009@lzu.cn}
        }
\affiliation{
    Institute of Theoretical Physics, Lanzhou University,
           Lanzhou 730000, People¡¯s Republic of China}

\begin{abstract}
With the new physical interpretation of quasinormal modes proposed
by Maggiore, the quantum area spectra of black holes have been
investigated recently. Adopting the modified Hod's treatment,
results show that the area spectra for black holes are equally
spaced and the spacings are in a unified form, $\triangle A=8\pi
\hbar$, in Einstein gravity. On the other hand, following
Kunstatter's method, the studies show that the area spectrum for a
nonrotating black hole with no charge is equidistant. And for a
rotating (or charged) black hole, it is also equidistant and
independent of the angular momentum $J$ (or charge $q$) when the
black hole is far from the extremal case. In this paper, we mainly
deal with the area spectrum of the stringy charged
Garfinkle-Horowitz-Strominger black hole, originating from effective
action that emerges in the low-energy string theory. We find that
both methods give the same results-that the area spectrum is equally
spaced and does not depend on the charge $q$. Our study may provide
new insights into understanding the area spectrum and entropy
spectrum for stringy black holes.
\end{abstract}


\date{\today}

\maketitle

\section{Introduction}
\label{secIntroduction}

The quantization of the black hole horizon area has become a
fascinating subject since the famous Bekenstein conjecture
\cite{Bekenstein1} was presented. Regarding the black hole horizon
area as an adiabatic invariant, Bekenstein obtained an equidistant
area spectrum $A_{n}= \epsilon\hbar\cdot n (n=0,1,2,...)$. After the
obtainment of the equally spaced area spectrum, many attempts have
been made to derive the area spectrum and entropy spectrum directly
from the dynamical modes of the classical theory
\cite{Bekenstein1995plb,Louko1996prd,
Dolgov1997plb,Barvinsky2001plb,Kastrup1996plb}. However, the area
spectrum has been somewhat controversial. In
\cite{Bekenstein1995plb}, the area spectrum was given as
$A_{n}=4\ln(k) \hbar\cdot n \;\;(k=2,3,...)$, but other methods
\cite{Barvinsky1996plb,Maggioreprl2008,Ropotenko2009prd} produced
$A_{n}=8\pi \hbar\cdot n$. In particular, in loop quantum gravity,
the area spectrum is in the form of $A=8\pi \gamma\Sigma_{j}
\sqrt{j(j+1)}$, with $\gamma$ the Immirzi parameter and $j=0, 1/2,
3/2,...$ \cite{Rovelli1995npb,Ashtekar1997cqg,Tanaka2008}.

An important step in this direction was made by Hod ten years ago.
He suggested that the spacing $\epsilon\hbar$ of the area spectrum
can be determined by utilizing the quasinormal mode frequencies of
an oscillating black hole \cite{Hodprl1998}. On the other hand,
Kunstatter pointed out that, for a system with energy $E$ and
vibrational frequency $\Delta \omega(E)$, the ratio $\frac{E}{\Delta
\omega(E)}$ is a natural adiabatic invariant
\cite{Kunstatterprl2003}. Interpreting the vibrational frequency
$\Delta \omega(E)$ as the real part of the quasinormal mode
frequencies and replacing the energy $E$ with the mass $M$, the area
spectrum of the Schwarzschild black hole was calculated as $A_{n} =
4\hbar\ln 3\cdot n $, which is consistent with Hod's result. This
rejuvenates a great interest in the investigation of the black hole
area spectrum via the interpretation of the quasinormal mode
frequencies
\cite{Hodprl1998,Dreyerprl2003,Polychronakosprd2004,Setarecqg2004,
Setareprd2004,Lepeplb2005,Vagenasjhep2008,Medvedcqg2008,
Kothawalaprd2008,Liplb2009,Cameliaplb2009,Medved2009,
Ansarinpb2008,Fernandoprd2009,Freidelcqg2008,Ortegaplb2009,
Majhi20099,Jingcpl2008,Kiselev2005,Kwon2010,Setare2010,
Medved22004,Hod22005cqg,Hod22007cqg,wei2009}.

Recently, Maggiore suggested that, in the high damping limit, the
proper frequency of the equivalent harmonic oscillator, which is
interpreted as the quasinormal mode frequency $\omega(E)$, should be
of the form \cite{Maggioreprl2008}
\begin{eqnarray}
\omega(E)=\sqrt{|\omega_{R}|^{2}+|\omega_{I}|^{2}}.
\end{eqnarray}
This form of proper frequency for the quasinormal mode frequency was
first pointed out in \cite{Wang2004prd}. Obviously, when $\omega_{I}
\ll \omega_{R}$, one could get $\omega(E)=\omega_{R}$ approximately
(adopted in the unmodified Hod's treatment), which has been extended
to other black holes \cite{Hodprl1998,Setarecqg2004,Dreyerprl2003}.
However, at highly excited quasinormal modes (i.e., $\omega_{R}\ll
\omega_{I}$), it is natural to get $\omega(E)= |\omega_{I}|$. With
the choice that the vibrational frequency
$\Delta\omega(E)=|\omega_{I}|_{n}-|\omega_{I}|_{n-1}$, the area
spectrum of the Kerr black hole was obtained by Vagenas
\cite{Vagenasjhep2008} with the modified Hod's treatment and
Kunstatter's method, respectively. The area spectrum calculated with
the modified Hod's treatment is equally spaced, while it is
nonequidistant and depends on the angular momentum parameter $J$
when employing Kunstatter's method. The methods disagree with each
other in describing the Kerr black hole. At the same time, Medved
also pointed out the disagreement between these two methods
\cite{Medvedcqg2008}. He argued that the quantum number $n$
appearing in the Bohr-Sommerfeld quantization condition is actually
a measure of the areal deviation from extremality. So, the
calculation of Kunstatter's method is restricted to small value of
$J$, which means that the black hole is far from the extremal black
hole, i.e., $M^{2}>>J$. In the spirit of this idea, the two methods
are found to coincide with each other and both give an equally
spaced and angular momentum parameter independent area spectrum when
the black hole is far from the extremal case. For other gravity
theories, for example, five-dimensional Gauss-Bonnet (GB) gravity,
both methods will give the result that the area spectrum is
nonequidistant \cite{Kothawalaprd2008,Weijhep2008,Banerjee2009}, but
the entropy spectrum is equidistant as first pointed out by
Kothawala et.al. \cite{Kothawalaprd2008}. However, if one sets the
GB coupling parameter $\alpha_{GB}\rightarrow 0$, the area spectrum
would be equidistant. A similar conclusion can also be found in
\cite{Majhi20099,Setare2010} for Ho\v{r}ava-Lifshitz gravity. So, it
is natural to make the following conclusions (from the new physical
interpretation of quasinormal modes proposed by Maggiore): (1) For a
nonrotating black hole with no charge, both the modified Hod's
treatment and Kunstatter's method can reproduce an equally spaced
area spectrum. (2) For rotating black holes (i.e. the Kerr black
hole and three-dimensional spinning black hole), the two methods
agree with each other when the black holes are far from the extremal
case and an equally spaced area spectrum can be reproduced. (3) For
other non-Einstein gravity theories, the two methods meet each other
and give the same result-that the area spectrum is not equally
spaced (details can been seen in
\cite{Kothawalaprd2008,Weijhep2008,Banerjee2009,Majhi20099,Setare2010}).

Although much work has been done and many issues have been clarified
in this field, another type of black hole solutions, obtained from
string theory, has not been discussed. These black holes have some
properties which are very different from the vacuum solutions
obtained from the frame of Einstein gravity. Take the
Garfinkle-Horowitz-Strominger (GHS) black hole, for example: its
temperature is the same as the Schwarzschild black hole. However, it
has charge-dependent area and entropy. Compared to the
Reissner-Nordstrom (RN) black hole, the stringy charged black hole
exhibits several different properties. First, it has only one
horizon, while the RN black hole has two. Second, the extremal cases
for this black holes are different. They occur at
$q^{2}=2M^{2}e^{2\phi_{0}}$ for the charged GHS black hole and
$q^{2}=M^{2}$ for the RN black hole. Based on these differences, it
is therefore worthwhile to investigate the area spectrum and entropy
spectrum for the stringy black hole. In this paper, we will study
these spectra for the charged GHS black hole. Employing the new
interpretation of the quasinormal modes, we obtain the area spectrum
and entropy spectrum for the charged GHS black hole using the
modified Hod's treatment and the Kunstatter's method, respectively.
We find that the two methods reproduce the same spacing of the area
spectrum for the GHS black hole. The area spectrum and entropy
spectrum are also found to be independent of the charge $q$. Our
results are also in agreement with the universal spacing in
\cite{Ropotenko2009prd}, where the author proposed an alternative
method for calculating the area spacing and demonstrated that this
method works not only for the Schwarzschild black hole, but also for
other charged and/or rotating black holes.

The paper is organized as follows. In Sec. \ref{Thermodynamic}, we
briefly review the thermodynamics of the charged GHS black hole and
show that the Bekenstein-Hawking entropy/area law holds. In Sec.
\ref{frequencies and spectrum}, with the new interpretation of the
quasinormal modes, we calculate the area spectrum and entropy
spectrum of the charged GHS black hole with the modified Hod's
treatment and Kunstatter's method, respectively. Finally, the paper
ends with a brief summary.

\section{Thermodynamic properties of the charged GHS black hole}
\label{Thermodynamic}

The GHS dilaton black hole is depicted by the following
four-dimensional low-energy action obtained from string theory:
\begin{eqnarray}
 I=\int d^{4}x \sqrt{-g}\bigg(-R-2(\nabla \phi)^{2}
 +e^{-2\phi}\mathcal {F}^{2}\bigg), \label{action}
\end{eqnarray}
where $\phi$ is a dilaton field and the Maxwell field $\mathcal
{F}_{\mu\nu}$ is associated with a $U(1)$ subgroup of $E_{8}\times
E_{8}$ or $Spin(32)/Z_{2}$. The charged black hole solution is given
by \cite{Garfinkle1991prd}
\begin{eqnarray}
 &&ds^{2}=-\bigg(1-\frac{2M}{r}\bigg)dt^{2}+\bigg(1-\frac{2M}{r}\bigg)^{-1}dr^{2}
          +r\bigg(r-\frac{q^{2}e^{-2\phi_{0}}}{M}\bigg)d\Omega^{2},\label{metric}\\
 &&e^{-2\phi}=e^{-2\phi_{0}}\bigg(1-\frac{q^{2}e^{-2\phi_{0}}}{M r}\bigg),\\
 &&\mathcal {F}=q\sin\theta d\theta\wedge d\varphi,
\end{eqnarray}
where $M$ and $q$ are related to the mass and charge of the black
hole, respectively. $\phi_{0}$ is the asymptotic constant value of
$\phi$ at $r\rightarrow \infty$. The metric (\ref{metric}) will
become the Schwarzschild metric as the charge $q\rightarrow 0$. The
radius of the event horizon is determined by
$g_{tt}=\frac{1}{g_{rr}}=0$, which gives
\begin{eqnarray}
 r_{+}&=&2M.  \label{horizon}
\end{eqnarray}
This result is consistent with that of the Schwarzschild black hole.
The area of the event horizon is calculated as
\begin{eqnarray}
 A    &=&\int \sqrt{g_{\theta\theta}g_{\varphi\varphi}}d\theta d\varphi \nonumber\\
       &=&4\pi r_{+}\bigg(r_{+}-\frac{q^{2}e^{-2\phi_{0}}}{M}\bigg)\nonumber\\
       &=&4\pi r_{+}^{2}-8\pi q^{2}e^{-2\phi_{0}}. \label{area}
\end{eqnarray}
Note that the area goes to zero at
$r_{+}=\frac{q^{2}e^{-2\phi_{0}}}{M}$ or
$q^{2}=2M^{2}e^{2\phi_{0}}$, which is related to the extremal black
hole. The Hawking temperature of the GHS black hole is similar to
the Schwarzschild case with $T=\frac{1}{8\pi M}$. The electric
potential computed on the horizon of the black hole is
\begin{eqnarray}
 V_{+}=\frac{q}{r_{+}}e^{-2\phi_{0}}.\label{charge}
\end{eqnarray}
Employing the first law of black hole thermodynamics, the entropy is
calculated as
\begin{eqnarray}
 S&=&\int \frac{dM-V_{+}dq}{T}\nonumber\\
  &=&4\pi M^{2}-2\pi q^{2}e^{-2\phi_{0}}+S_{0}, \label{entropy2}
\end{eqnarray}
where $S_{0}$ is an integral constant. Substituting Eqs.
(\ref{horizon}) and (\ref{charge}) into Eq. (\ref{entropy2}), we
obtain the entropy/area law
\begin{eqnarray}
 S=\frac{A}{4}+S_{0}. \label{entropy}
\end{eqnarray}
Setting the constant $S_{0}\rightarrow 0$, the result exactly
confirms the standard Bekenstein-Hawking entropy/area law. In other
word, if the relation (\ref{entropy}) holds, the first law of black
hole thermodynamics will be satisfied naturally. For the fixed
charge, the heat capacity is expressed as
\begin{eqnarray}
 C_{q}=\bigg(\frac{\partial M}{\partial T}\bigg)_{q}=-8 \pi M^{2}.
\end{eqnarray}
This negative heat capacity implies that this black hole could not
stably exist in a heat bath.

\section{Quasinormal frequencies and area spectrum}
\label{frequencies and spectrum}

Introducing a coordinate transformation $\rho=\sqrt{r(r-2b)}$, the
metric (\ref{metric}) can be expressed as
\begin{eqnarray}
 &&ds^{2}=-\bigg(1-\frac{2M}{b+\sqrt{b^{2}+\rho^{2}}}\bigg)dt^{2}
          +\bigg(1-\frac{2M}{b+\sqrt{b^{2}+\rho^{2}}}\bigg)^{-1}
              \frac{\rho^{2}}{b^{2}+\rho^{2}}d\rho^{2}
          +\rho^{2}d\Omega^{2},
\end{eqnarray}
where $b={q^{2}e^{-2\phi_{0}}}/{(2M)}$. The event horizon of the
black hole now locates at
\begin{eqnarray}
 \rho_{+}=2\sqrt{M^2-\frac{1}{2}{q^{2}e^{-2\phi_{0}}}},
\end{eqnarray}
while the Hawking temperature is still given by $T=\frac{1}{8\pi
M}$.

In the dilaton spacetime, the general perturbation equation for a
coupled massless scalar field is given by
\begin{eqnarray}
 \nabla^{2}\Phi-\zeta R \Phi=0, \label{equation}
\end{eqnarray}
where $\Phi$ is the scalar field and $R$ is the Ricci scalar
curvature. The parameter $\zeta$ denotes the coupling between the scalar field
and the gravitational field. The Laplace-Beltrami operator is
\begin{eqnarray}
 \nabla^{2}=\frac{1}{\sqrt{-g}}\partial_{\mu}
 (\sqrt{-g}g^{\mu\nu}\partial_{\nu}).
\end{eqnarray}
Adopting the WKB approximation $\Phi={e^{-i\omega
t}}{r^{-1}}f(r)Y(\theta,\varphi)$ and imposing the proper boundary
conditions, the asymptotic quasinormal frequencies were obtained in
\cite{Chen2005cqg}:
\begin{eqnarray}
 \frac{\omega}{T}=\ln(1+2\cos(\sqrt{2\zeta}\pi))+i(2k+1)\pi,\quad k\rightarrow
 \infty. \label{quasinormal}
\end{eqnarray}
It is worth to noting that the quasinormal frequencies not only
depend on the parameters of the black hole, but also on the coupling
constant $\zeta$. Ignoring the second term in Eq. (\ref{equation})
or taking the limit $\zeta\rightarrow0$, the real part of the
quasinormal frequencies will become $T\ln 3$, which is consistent
with the Schwarzschild case. For other black holes, the quasinormal
modes were studied in
\cite{Wang2000plb,Wang2001prd,Wang2002prd,Wang2004prdd,Wang2001prdd}.

Now, we will study the area spectrum of the charged GHS black hole
via the quasinormal frequencies (\ref{quasinormal}) and try to find
out whether the area spectrum is equidistant. First, we will
consider the area spectrum of the GHS black hole with the modified
Hod's treatment, where the variations of the black hole parameters
are regarded as finite quantities. Thus, the variations of the
parameters satisfy
\begin{eqnarray}
 \Delta M-V_{+}\Delta q=\hbar\Delta \omega(E). \label{ww}
\end{eqnarray}
With the interpretation of identifying the vibrational frequency
$\Delta\omega(E)$ as the real part of the quasinormal modes, much
work have been done (e.g.,
\cite{Hodprl1998,Setarecqg2004,Dreyerprl2003}). However, as proposed
by Maggiore, when $\omega_{R}\ll \omega_{I}$, we should take
$\omega(E)=|\omega_{I}|$, approximately. Following this choice, the
$\Delta \omega$ that appeared in (\ref{ww}) is calculated as
\begin{eqnarray}
 \Delta\omega=|\omega_{I}|_{k}-|\omega_{I}|_{k-1}=2\pi T.\quad (k\gg 1) \label{wen}
\end{eqnarray}
From the formulas (\ref{horizon}), (\ref{charge}) and (\ref{wen}),
we get
\begin{eqnarray}
 4M\Delta M-2qe^{-2\phi_{0}}\Delta q=\hbar.
\end{eqnarray}
Recalling the area (\ref{area}), it is easy to obtain the spacing of
the area spectrum
\begin{eqnarray}
 \Delta A=8\pi \hbar. \label{spectrum}
\end{eqnarray}
So far, we have obtain a charge-independent spacing of the area
spectrum for the charged GHS black hole which is in full agreement
with that of the Schwarzschild black hole given by Maggiore. It
might be worth mentioning that the black hole here is not restricted
to the case of a far-from-extremal black hole. However, we will see
in the following that the black hole must be a far-from-extremal one
in the calculation using Kunstatter's method.

Next, following Kunstatter's method, we would like to reconsider the
area spectrum of the charged GHS black hole. Here, we regard the
variations of the black hole parameters as infinite quantities,
which can ensure that the formula could be written in an integral
form. Given a system with energy $E$ and vibrational frequency
$\Delta\omega(E)$, a natural adiabatic invariant quantity proposed
by Kunstatter \cite{Kunstatterprl2003} is
\begin{eqnarray}
 I=\int \frac{dE}{\Delta \omega(E)}. \label{condition}
\end{eqnarray}
At the large $n$ limit, the relation between the adiabatic invariant
quantity $I$ and Bohr-Sommerfeld quantization is of the form
\begin{eqnarray}
 I\approx n\hbar, \quad n\rightarrow \infty \label{Bohr}
\end{eqnarray}
Here we need to note that the application of the Bohr-Sommerfeld
quantization condition requires $n$ to be a very large number, which
also implies that the black hole must be far away from extremality,
as the number $n$ is a measure of the areal deviation from
extremality. In general, the adiabatic invariant quantity $I$ of a
black hole can be expressed as
\begin{equation}
 I=\int \frac{dE}{\Delta\omega}
  =\int \frac{dM-\Omega dJ-V_{+}dq}{\Delta\omega}, \label{adiabatic}
\end{equation}
where $\Omega$ and $V_{+}$ are the angular velocity and the electric
potential on the horizon, respectively. In the second step, the
first law of black hole thermodynamics is used. For the charged GHS
black hole, the adiabatic invariant quantity (\ref{adiabatic})
reduces to
\begin{equation}
 I=\int \frac{dE}{\Delta\omega}
  =\int \frac{dM-V_{+}dq}{\Delta\omega}. \label{2adiabatic}
\end{equation}
The vibrational frequency $\Delta\omega$ is given in (\ref{wen}).
Substituting $\Delta\omega$ into (\ref{2adiabatic}), we obtain the
adiabatic invariant quantity
\begin{eqnarray}
 I
   &=&\int 4MdM-\int\frac{4Mqe^{-2\phi_{0}}}{r_{+}}dq
   \nonumber\\
   &=& \frac{1}{2}r_{+}^{2}-q^{2}e^{-2\phi_{0}}.
\end{eqnarray}
Using the Bohr-Sommerfeld quantization condition (\ref{Bohr}), at
the large $n$ limit, we obtain
\begin{equation}
 2M^{2}-q^{2}e^{-2\phi_{0}}=n \hbar.
\end{equation}
Recalling the area from (\ref{area}), the area spectrum of this
black hole is given by
\begin{equation}
 A_{n}=8\pi\hbar \cdot n, \label{BTZresult}
\end{equation}
with the spacing $\Delta A=A_{n}-A_{n-1}=8\pi\hbar$. This spacing is
consistent with that of (\ref{spectrum}),
which is equally spaced. It is also obvious that the spacing of the
area spectrum is independent of the charge $q$ of the black hole.
Recalling the relationship (\ref{entropy}), one could get the
entropy spectrum
\begin{equation}
 S_{n}=2\pi\hbar \cdot n+S_{0},
\end{equation}
with the spacing
\begin{equation}
 \Delta S=S_{n+1}-S_{n}=2\pi\hbar.
\end{equation}
The entropy is also equidistant and the spacing is independent of
the charge $q$ of the black hole. On the other hand, if one
interprets the vibrational frequency $\Delta \omega(E)$ as the real
part of the quasinormal modes, i.e. $\Delta
\omega(E)=T\ln(1+2\cos(\sqrt{2\zeta}\pi))$, one could obtain the
area spectrum $A_{n}=4\ln(1+2\cos(\sqrt{2\zeta}\pi))\hbar\cdot n$.
Taking the coupling $\zeta=0$, the spectrum is
$A_{n}=4\ln(3)\hbar\cdot n$, which is the same as in Schwarzschild
black hole.

We are almost finished with the calculations of this paper. For
completeness, we would like to give a brief review on the study of
the area spectra for the charged black holes. A reasonable
discussion about our results and other charged black holes is also
given.

In \cite{Setareprd2004}, Setare studied the area spectrum of the
four-dimensional extreme RN black hole by regarding that the real
part of the quasinormal frequency for the extreme RN black hole is
the same as that of the Schwarzschild black hole. The area spectrum
and entropy spectrum are given as $A_{n}=\ln 3\;\hbar \cdot n$ and
$S_{n}=\frac{\ln 3}{4}\hbar \cdot n$, respectively. The area and
entropy spectra are both equally spaced. However the values of the
spacings are not consistent with our results. For the nonextreme RN
black hole, Hod showed the equally spacings $\Delta A=4\hbar \ln 2$
and $\Delta A=4\hbar \ln 3$ with two distinct families of
quasinormal frequencies \cite{Hod22005cqg,Hod22007cqg}.

Using the reduced phase-space quantization, Barvinsky et.al.
\cite{Barvinsky2001plb} also calculated the area spectrum for the
charged black hole. For the four-dimensional RN black hole, they
found the area spectrum $A_{n,p}=4\pi \hbar(2n+p+1)$ with two
quantum numbers $n, p=0,1,2,....$, where the charge quantum number
$p$ corresponds to $Q=\pm\sqrt{\hbar p}$. Although this area
spectrum is quantized, it is charge-dependent as the charge quantum
number $p$ comes into the spectrum.

After this paper was completed, some results about the area spectrum
for the $d$-dimensional RN black hole were obtained in
\cite{Lopez2010}. For the black hole far from the extremal case
(small charge limit), it is found that the area spectrum is equally
spaced with $A_{n}=8\pi \hbar\cdot n$, following Kunstatter's
method.

Our results show that the modified Hod's treatment and Kunstatter's
method give a unified description of the area spectrum of the
charged GHS black hole. The area spectrum and entropy spectrum are
found to be equally spaced, in agreement with the universal one. The
spacing is also found to be charge-independent, so there are some
differences from that of \cite{Barvinsky2001plb}. However, this
quasinormal frequencies method is not easy to generalize to the
other charged and/or rotating black holes owing to the fact that
their quasinormal frequencies are generally not in a closed form
unless some approximations are taken. However, we can claim that,
the two methods work for the charged GHS black hole, and the equally
spaced area spectrum and entropy spectrum are obtained. We believe
that our study into the area spectrum of the charge GHS black hole
should provide new insights on understanding the area and entropy
spectra for other stringy black holes.

\section{Summary}
\label{Conclusion}

In this paper, we mainly deal with the area spectrum and entropy
spectrum of a stringy charged GHS black hole, which originate from
the effective action that emerges in low-energy string theory. The
black hole have some properties that are very different from the
vacuum solutions obtained from the frame of Einstein gravity. With
the new physical interpretation of quasinormal modes, the area
spectrum and entropy spectrum for the black hole are obtained
following the modified Hod's treatment and the Kunstatter's method,
respectively. Despite the different properties between the black
holes in the low-energy string theory and in Einstein gravity, the
area spectrum and the entropy spectrum of the black hole are found
to be equally spaced, which is consistent with the results for the
charged RN black hole \cite{Lopez2010}. Furthermore, these spectra
are also found to be independent of the charge $q$, which is
different from the results of
\cite{Setareprd2004,Hod22005cqg,Hod22007cqg,Barvinsky2001plb},
obtained with a different physical interpretation of quasinormal
modes or other methods. This quasinormal modes method is hard to
extend to other charged and/or rotating black holes, as the
quasinormal modes generally can not be obtained in a closed form.
However, we believe that our study on the charged GHS black hole
should provide new insights into understanding the area spectrum and
entropy spectrum for other stringy black holes.

\section*{Acknowledgements}

This work was supported by the Program for New Century Excellent
Talents in University, the National Natural Science Foundation of
China (No. 10705013), the Doctoral Program Foundation of
Institutions of Higher Education of China (No. 20070730055 and No.
20090211110028), the Key Project of Chinese Ministry of Education
(No. 109153), the Natural Science Foundation of Gansu Province,
China (No. 096RJZA055), and the Fundamental Research Funds for the
Central Universities (No. lzujbky-2009-54 and No. lzujbky-2009-163).

\end{document}